\documentclass[conference]{IEEEtran}
\IEEEoverridecommandlockouts
\usepackage{cite}
\usepackage{amsmath,amssymb,amsfonts}
\usepackage{algorithmic}
\usepackage{graphicx}
\usepackage{textcomp}
\usepackage{xcolor}
\usepackage{graphicx}
\usepackage{booktabs}
\usepackage{booktabs}
\usepackage{diagbox}
\usepackage{svg}

\def\BibTeX{{\rm B\kern-.05em{\sc i\kern-.025em b}\kern-.08em
    T\kern-.1667em\lower.7ex\hbox{E}\kern-.125emX}}
\begin{document}

\title{Vibe Coding on Trial: Operating Characteristics of Unanimous LLM Juries\\

}

\author{%
\IEEEauthorblockN{Muhammad Aziz Ullah, Abdul Serwadda}
\IEEEauthorblockA{Texas Tech University, Lubbock, Texas\\
Email: mazizull@ttu.edu, abdul.serwadda@ttu.edu}%
}
 
\maketitle

\begin{abstract}
Large Language Models (LLMs) are now good enough at coding that developers can describe intent in plain language and let the tool produce the first code draft, a workflow increasingly built into tools like GitHub Copilot, Cursor, and Replit. What is missing is a reliable way to tell which model written queries are safe to accept without sending everything to a human. We study the application of an LLM jury to run this review step. We first benchmark 15 open models on 82 MySQL text to SQL tasks using an execution grounded protocol to get a clean baseline of which models are strong. From the six best models we build unanimous committees of sizes 1 through 6 that see the prompt, schema, and candidate SQL and accept it only when every member says it is correct. This rule matches safety first deployments where false accepts are more costly than false rejects. We measure true positive rate, false positive rate and Youden J and we also look at committees per generator. Our results show that single model judges are uneven, that small unanimous committees of strong models can cut false accepts while still passing many good queries, and that the exact committee composition matters significantly.
\end{abstract}

\section{Introduction}
Large Language Models (LLMs) have become good enough at writing code that many developers now describe intent in natural language and let the model produce a first draft of the code. This natural-language–to–code interaction is no longer limited to general-purpose chatbots; it is being built directly into development tools such as GitHub Copilot, Cursor, and Replit ~\cite{github_copilot,cursor_ai,replit_ai}. In practitioner circles this workflow is widely referred to as ``vibe coding''~\cite{karpathy_vibe_coding,heath_vibe_coding}. It accelerates scaffolding, reduces boilerplate, and shortens iteration. The hard question is when an LLM-written answer is safe to accept. A snippet can execute and still be wrong. A patch can pass a narrow test and still mishandle edge cases. Two models can disagree confidently about the very same output.

Human review helps, but it does not scale. Modern teams ship large volumes of small changes across many services and time zones. Expert reviewers are scarce and expensive, review queues create latency, repetitive checks are error prone, and context switching erodes quality. Organizations need an automated first line of acceptance so that human effort is reserved for the minority of cases that truly require it.

Most evaluations still treat the model purely as a coder, asking only whether its output matches a reference or passes a test. Real deployments also need a critic that can say, with measurable reliability, ``this looks correct.'' Prior work has explored using a single LLM as a judge, but single-judge setups inherit that model's biases and provide little guidance on system-level operating behavior.

In this paper we treat review as a system design problem and build committees of models that act as a jury over AI-written code. We begin by benchmarking \textbf{15} of the strongest open-source LLMs as \emph{coders} on \textbf{82} natural-language prompts that span practical SQL needs: simple selections and filters, multi-table joins, grouping and aggregation (including \texttt{HAVING} and \texttt{NULL} semantics), nested subqueries, and cases that are easy to get subtly wrong (for example, duplicates versus \texttt{COUNT(DISTINCT)}). All experiments target \textbf{MySQL}, a widely used open-source DBMS. 

For each prompt we compare a model's first SQL answer to a human-authored ground-truth query using an execution-based check on independently generated database instances. This yields a clean single-attempt accuracy baseline and, to our knowledge, one of the largest open-only text-to-SQL coder comparisons at this scale.

Leveraging this baseline, we select the \textbf{top six} models by execution accuracy to form the \emph{judge pool}. While a weaker generator could in principle be a competent judge, we restrict the pool to strong coders for a pragmatic reason: teams are more likely to trust reviewers that also demonstrate strong coding performance.

From this shortlisted pool we instantiate sub-committees of sizes \(N \in \{1,2,3,4,5,6\}\) (that is, all \(\binom{6}{k}\) subsets for each \(k\)). Each committee member evaluates a candidate SQL query given the prompt and schema and returns a binary judgment (correct or incorrect). To consolidate votes we adopt a \textbf{unanimous} (AND) rule: a candidate is accepted only if \emph{every} committee member votes ``correct.'' This intentionally conservative choice targets \emph{safe acceptance} for production settings where a false accept can silently corrupt analyses or downstream decisions. Unanimity is better aligned with the ''vibe coding'' use case than majority voting because the loss is asymmetric (false accepts are costlier than false rejects that fall back to human review), errors across judges can be correlated (and simple majorities can amplify a shared bias), and the rule is operationally simple and auditable.

We summarize how committee behavior evolves with size and composition, including how often committees accept, how strict they are in practice, and how their decisions vary across sources of candidate code. Beyond aggregate views, we analyze \emph{committee\(\times\)generator} slices to reveal which committees best protect against specific generators' failure modes and where the costs of conservatism (e.g., lower acceptance) outweigh its benefits. This framing is general to code in any language; we validate it in text-to-SQL because SQL offers crisp semantics and execution-grounded checks that make committee judgments auditable and reproducible.

The paper makes the following \textbf{contributions.}
\begin{itemize}
    \item \textbf{A contemporary open-model baseline for text-to-SQL.}
    We benchmark \emph{15} strong open-source LLMs as coders on \emph{82} prompts that span practical SQL needs (from simple selections to multi-join, aggregation, and nested queries) on \textbf{MySQL}. 
    Each model is scored by single-attempt, execution-based comparison to a human-authored ground truth across independently generated databases. 
    This establishes a reproducible, open-only baseline at a scale that, to our knowledge, has not been reported for SQL "vibe coding."

    \item \textbf{Conservative committees of strong judges.}
    Using the coder baseline, we select the \emph{top six} models as a judge pool and instantiate all committees of sizes \(N\in\{1,2,3,4,5,6\}\). 
    Each committee applies a \emph{unanimous} acceptance rule (AND): a candidate query is accepted only when every member votes “correct.” 
    We characterize how acceptance behavior evolves with committee size and composition, including the practical trade-off between safety and throughput.

    \item \textbf{Committee behavior per SQL generator.}
    Beyond aggregate reporting, we analyze \emph{committee\(\times\)generator} outcomes: how each committee judges outputs from each coder. 
    These slices reveal complementary and redundant judge combinations, identify committees that best protect against specific generators’ failure modes, and highlight where conservatism delivers diminishing returns.
\end{itemize}

\section{Related Research}
We organize prior work into three thrusts: (i) studies that evaluate \emph{LLMs as coders} by comparing generated programs to ground truth with unit tests and benchmark suites; (ii) work that uses \emph{LLMs as critics}, where a single model judges or ranks outputs from other models; and (iii) \emph{committee-based evaluation}, which aggregates multiple models to improve reliability. 

\subsection{LLMs as Coders}
Large language models are commonly evaluated on programming benchmarks that compare generated code to ground truth via unit tests. Chen et al.\ introduced Codex alongside HumanEval and the now-standard \emph{pass@k} protocol \cite{chen2021codex}. MBPP extends this style to a larger set of mostly basic Python problems \cite{austin2021}. These protocols have become the de facto baseline for coder LLMs across both open and proprietary releases.

% Beyond unit-test suites, state-of-the-art coders are evaluated in \emph{competitive programming} and other real-world settings. AlphaCode demonstrated competition-level performance via massive sampling and execution-based filtering \cite{li2022alphacode}. OpenAI’s recent large reasoning models (e.g., \emph{o1}, \emph{o1-ioi}, early \emph{o3}) report strong results on algorithmic coding tasks and live contests under controlled protocols \cite{elkishky2025o1ioi,codeforces_o1_blog}, and DeepSeek-R1 achieves expert-level performance on Codeforces \cite{guo2025deepseek}. In parallel, newer releases emphasize broader software-engineering benchmarks: GPT-5 reports state-of-the-art results on SWE-bench Verified \cite{openai_gpt5}, while xAI’s Grok-2 reports gains across reasoning and coding evaluations \cite{xai_grok2}.

State-of-the-art models are also tested on \emph{competitive programming} and real-world tasks. AlphaCode demonstrated competition-level performance through massive sampling and filtering \cite{li2022alphacode}. OpenAI's reasoning models (\emph{o1}, \emph{o1-ioi}, early \emph{o3}) show strong results on algorithmic contests \cite{elkishky2025o1ioi,codeforces_o1_blog}, and DeepSeek-R1 reaches expert-level Codeforces performance \cite{guo2025deepseek}. Recent releases also target software-engineering benchmarks: GPT-5 achieves state-of-the-art on SWE-bench Verified \cite{openai_gpt5}, while Grok-2 shows gains across reasoning and coding tasks \cite{xai_grok2}.

For text-to-SQL in particular, Zhang et al.\ benchmark a broad slate of LLMs (open and proprietary) across multiple text-to-SQL sub-tasks—generation, debugging, optimization, and schema linking—under a unified, evaluation-based protocol \cite{zhang2024t2sql_benchmark}. Liu et al.\ evaluate ChatGPT in a comprehensive zero-shot setting across twelve datasets, showing strong baselines without fine-tuning \cite{liu2023chatgpt_t2sql}. Pourreza and Rafiei improve LLM text-to-SQL via decomposed in-context prompting with self-correction, yielding substantial execution-accuracy gains \cite{pourreza2023din}.

\noindent\textbf{How we differ.}
These works treat the model as a \emph{coder} and score generation accuracy. We instead study \emph{judgment}: committees of models deciding whether a candidate SQL query should be accepted, grounded by execution-based checks.

\subsection{LLMs as Critics}
To reduce the cost and variability of human evaluation, many studies use \textit{LLM-as-a-judge}, where a strong model is prompted to rate or rank other models’ outputs. This appears across proprietary judges (GPT-4/4o, Claude, Gemini) and strong open judges based on Llama-family checkpoints. For example, Zheng et al.\ introduce MT-Bench and use an automated judge in Chatbot Arena for scalable dialogue assessment \cite{zheng2023judging}; Liu et al.\ propose G-Eval and show that prompted judges can correlate well with human preferences on summarization and dialogue \cite{liu-etal-2023-g}. Such single-judge setups are now common in public leaderboards and internal evaluations.

\noindent\textbf{How we differ.}
Single-judge evaluation inherits one model’s biases and provides limited guidance on operating behavior. We move from a single arbiter to \emph{committees} of strong (open) judges and use a \emph{unanimous} decision rule that matches safety-first deployments: accept only when every judge agrees. Rather than preference scores, our labels are execution-grounded on SQL, and we report how acceptance behavior changes with committee size and which committees protect best against particular generators’ failures.

\subsection{Committee-Based Evaluation}
Using multiple models as an ensemble or “committee” connects to long-standing ideas in ML and software engineering. In classical ensemble learning, aggregating diverse predictors improves robustness; in software fault tolerance, \emph{N}-version programming runs independently implemented modules in parallel and aggregates their outputs to tolerate individual failures \cite{avizienis1985nversion}. 

Recent LLM work explores multi-agent evaluators, largely on \emph{non-code} tasks. Chan et al.\ (\textsc{ChatEval}) use role-based referees that debate and assess open-ended generations, improving agreement with human judgments over a single judge \cite{chan2024chateval}. Kumar et al.\ (\textsc{CourtEval}) adopt a courtroom metaphor (grader, critic, defender) to calibrate summarization and dialogue evaluations \cite{kumar-etal-2025-courteval}. Zhao et al.\ (\textsc{Language Model Council}) aggregate up to 20 models that create questions and vote on answers, yielding rankings that better track human preferences \cite{zhao-etal-2025-language}. Earlier “debate’’ proposals similarly target general reasoning with human adjudicators, not code execution signals \cite{irving2018debate}.

\noindent\textbf{How we differ.}
Prior committees typically use majority-like rules on subjective text (not SQL code) quality and report a single operating point. We target \emph{program correctness} with \emph{execution-grounded} labels in text-to-SQL, and we adopt the \emph{unanimous} rule to prioritize safe acceptance in production. We instantiate all committees from a top-six pool of strong open models, summarize how acceptance behavior evolves with committee size and composition, and analyze committee\(\times\)generator outcomes to identify complementary versus redundant judge combinations. Our setup uses only open models and a transparent MySQL protocol, enabling replication and extension.

\begin{table}[ht]
\centering
\vspace{1em}
\caption{\textbf{Model Specifications}}
\label{tab:model_specs}
\begin{tabular}{|c|l|l|l|}
\hline
\textbf{Model ID} & \textbf{\shortstack{Model Name}} & \textbf{\shortstack{Parameters}}  & \textbf{Company} \\ \hline
\textbf{1} & gpt-oss:120b & 120B & OpenAI \\ \hline
\textbf{2} & gpt-oss:20b & 20B & OpenAI \\ \hline
\textbf{3} & codegemma:7b & 7B & Google \\ \hline
\textbf{4} & phi4:14b & 14B & Microsoft \\ \hline
\textbf{5} & gemma3:27b & 27B & Google \\ \hline
\textbf{6} & codellama:34b & 34B & Meta \\ \hline
\textbf{7} & llama3.1:70b & 70B & Meta \\ \hline
\textbf{8} & phind-codellama:34b & 34B & Phind \\ \hline
\textbf{9} & codellama:7b & 7B & Meta \\ \hline
\textbf{10} & llama3.1:8b & 8B& Meta \\ \hline
\textbf{11} & wizardcoder& 7B & WizardLM \\ \hline
\textbf{12} & gemma:7b & 7B & Google \\ \hline
\textbf{13} & mistral:7b & 7B & Mistral AI \\ \hline
\textbf{14} & sqlcoder:7b & 7B & Defog \\ \hline
\textbf{15} & falcon:7b & 7B & TII \\ \hline
\end{tabular}
\end{table}

\section{Experiment Design}
\subsection{Tools, Platforms, and Models Used}
\label{sec:tooling}
All experiments run within a unified Python evaluation pipeline that (i) queries language models, (ii) generates and executes SQL against fresh databases, and (iii) records outputs for our subsequent analysis. Our setup has the following components:

\paragraph{Ollama (model runtime).}
We use \textbf{Ollama} to host open models locally and exchange requests and responses through a single, consistent API. Decoding is fixed at \(\mathrm{temperature}=0\) to suppress sampling variability and make results reproducible across runs and committees. Each generation call provides the database schema and a brief instruction that asks the model to produce a \emph{MySQL} query answering the user’s question and to return \emph{only} the SQL (no explanation or extra text).%
\footnote{For judging, we similarly provide the prompt, schema, and a candidate SQL query and ask the model for a binary assessment of correctness.}
Table~\ref{tab:model_specs} lists the \emph{15} open models we serve via Ollama, including parameter count and proprietor/maintainer. These models are evaluated first as coders; the best of them then form the judge pool for committee experiments.

\paragraph{MySQL (database engine).}
We use \textbf{MySQL} as the database engine because it is the most widely used open-source DBMS and has stable semantics for the features we target (joins, aggregation/\texttt{HAVING}, nested queries, \texttt{NULL}). We pin the server to \textbf{MySQL 8.0.44} and enable \texttt{ONLY\_FULL\_GROUP\_BY} in \texttt{sql\_mode}, which enforces standard grouping semantics: every non-aggregated column referenced in \texttt{SELECT}/\texttt{HAVING}/\texttt{ORDER BY} must appear in \texttt{GROUP BY}, preventing ambiguous rows and non-deterministic results. Since many syntactically different statements are semantically equivalent, we judge correctness of SQL queries by query execution \emph{results}, not SQL query text; the full execution-grounded protocol is in Section \ref{methods-1}.

\paragraph{Python tooling (LangChain + SQLAlchemy).} We use \textbf{LangChain} only as a light wrapper to issue model calls through Ollama and to log prompts and responses; prompt templates, vote parsing, and committee logic are implemented in plain Python. \textbf{SQLAlchemy} manages connections to MySQL, query execution with timeouts, and result retrieval. The pipeline injects the schema into prompts, trims outputs to the first complete statement, and normalizes result sets before comparison.

\subsection{Database Schema and Queries}
\label{schema-1}

To systematically evaluate our approach, we construct a representative experimental environment centered on a relational database modeling an online retail store. This schema captures the essential data relationships found in real-world e-commerce systems while providing sufficient complexity to exercise diverse query patterns.

\subsubsection{Database Schema}
The experimental database comprises six interconnected tables that mirror typical e-commerce operations. The \texttt{Users} table maintains customer information including authentication credentials (username), contact details (email), demographic attributes (name, age, location), account metadata (creation timestamp), and activity status. The \texttt{Products} table catalogs merchandise with standard retail attributes: unique identifiers, descriptive information, pricing, inventory levels, and categorical classification via foreign key reference to \texttt{Categories}. This hierarchical organization enables product taxonomy through the \texttt{Categories} table, which maintains category identifiers, descriptive metadata, and temporal tracking of catalog additions.

Order management is handled through two related tables. The \texttt{Orders} table records order-level information including order identifier (primary key), user identifier (foreign key), total amount, order status (e.g., pending, complete), and shipping address. The \texttt{Order\_Items} table captures line-item details with its own identifier, references to both the parent \texttt{Orders} table and associated \texttt{Products} table (via foreign keys), and item quantity. Finally, the \texttt{Reviews} table enables customer feedback through a structure containing review identifier, user and product references (foreign keys), numerical rating, review text, and submission date.

\subsubsection{Query Dataset}
To comprehensively evaluate query performance, we developed a dataset of 82 carefully crafted SQL queries spanning diverse Data Manipulation Language (DML) operations and complexity levels. The queries are designed to reflect realistic analytical workloads in an e-commerce context, with systematic coverage of both fundamental and advanced SQL constructs.

\textbf{Query Composition and Complexity Distribution.}
The dataset encompasses five primary complexity tiers, each targeting distinct aspects of query processing and optimization.

\textit{Basic Retrieval:} Simple \texttt{SELECT} statements with optional filtering and sorting, including full table retrievals, single-column projections, \texttt{WHERE} clause predicates on various data types (Boolean, string, numeric, temporal), and \texttt{ORDER BY} operations.

\textit{Aggregate Analysis:} Single-table and cross-table aggregations utilizing \texttt{COUNT}, \texttt{SUM}, \texttt{AVG}, \texttt{MIN}, and \texttt{MAX} functions. This tier includes both standalone aggregations and those combined with \texttt{LEFT JOIN}s to handle null-preserving relationships, with results grouped by categorical dimensions.

\textit{Multi-Table Joins and Grouping:} Complex join patterns including \texttt{INNER}, \texttt{LEFT}, and \texttt{RIGHT} joins across 2--3 tables, often combined with \texttt{GROUP BY} clauses and aggregate functions. Queries incorporate temporal grouping functions (\texttt{YEAR}, \texttt{MONTH}) and multiple simultaneous aggregations on joined datasets.

\textit{Filtered Aggregations with HAVING:} Advanced analytical queries combining joins, aggregations, grouping, and post-aggregation filtering via \texttt{HAVING} clauses. This tier tests the optimizer's ability to efficiently process multi-stage analytical pipelines with various threshold conditions on aggregated metrics.

\textit{Subqueries and Advanced Constructs:} The most complex tier featuring scalar subqueries, correlated subqueries, nested subqueries with multiple levels of nesting, \texttt{IN}/\texttt{NOT IN} predicates with subqueries, and window functions (\texttt{ROW\_NUMBER() OVER} with \texttt{PARTITION BY}). Queries include comparative analytics (e.g., ``above average'' conditions), relational division patterns (universal quantification), temporal cohort analysis, \texttt{DISTINCT} operations with subqueries, and cumulative window aggregations (\texttt{SUM() OVER} with \texttt{ORDER BY}).

% The query set deliberately includes edge cases such as \texttt{NULL} handling, date arithmetic with various intervals (30, 60, 90 days, months, years), \texttt{LIMIT} clauses for top-$N$ queries, and combinations of multiple advanced features within single queries. This stratified design ensures comprehensive coverage of query optimizer decision spaces and execution engine capabilities, providing a robust foundation for performance analysis across diverse workload characteristics.

\subsection{Methodology and Metrics for LLM Evaluation}
\label{methods-1}
\subsubsection{Execution-Grounded Labeling Protocol}
To judge whether a query generated by a given LLM matches the ground-truth query, exact string match is not meaningful because multiple syntactic forms can be semantically equivalent. We therefore label correctness by comparing \emph{results}, not SQL query text. Our protocol for judging the correctness of a query is as follows: 

\begin{enumerate}
  \item \textbf{Instances.} Create \emph{ten} independently seeded databases under the schema in Section \ref{schema-1}.
  \item \textbf{Reference.} Execute the human-authored ground-truth query on each instance and record a \emph{canonicalized} result (stable row/column ordering, consistent \texttt{NULL} handling, fixed numeric precision).
  \item \textbf{Candidate.} Execute the candidate query under identical settings with a wall-clock timeout and a maximum row cap; canonicalize results in the same way.
  \item \textbf{Decision.} Label the candidate \textbf{correct} only if its result matches the ground truth on \emph{all ten} instances; otherwise label it \textbf{incorrect}.
\end{enumerate}
This multi-instance check reduces accidental agreement on any single dataset and yields an auditable, reproducible notion of correctness aligned with operational use.

\subsubsection{Metrics and Reporting}
\label{sec:metrics}
We characterize each committee’s behavior using three complementary metrics: \emph{true–positive rate} (TPR): the proportion of correct SQL queries that the committee judges to be correct, \emph{false–positive rate} (FPR): the proportion of wrong SQL queries that the committee judges to be correct, and \emph{Youden’s} \(J\) (defined as \(J=\mathrm{TPR}-\mathrm{FPR}\)).

\paragraph{Why these metrics.}
\textbf{FPR} captures \emph{risk}: the tendency to accept incorrect SQL (false accepts are costly in deployment).
\textbf{TPR} captures \emph{utility}: the ability to accept correct SQL (avoiding unnecessary human review).
\textbf{Youden’s \(J\)} summarizes \emph{discriminative value} by rewarding high TPR and penalizing FPR in a class imbalance-neutral way; unlike raw accuracy, \(J\) is not skewed by the mix of correct vs.\ incorrect items. Together, TPR, FPR, and \(J\) expose both sides of the safety–throughput trade-off and a single composite for ranking committees.

\paragraph{All committees at a given size.}
From the six-model judge pool, we enumerate every committee of size \(k \in \{1,2,3,4,5,6\}\) (i.e., all \(\binom{6}{k}\) subsets) and evaluate each under the \emph{unanimous} rule. We report two complementary views. In the \emph{aggregate view}, each committee is evaluated on all \(82 \times 6 = 492\) candidate queries pooled across generators; items are treated uniformly, and results summarize overall operating behavior at fixed \(k\). In the \emph{per-generator view}, we condition on the generator \(g\) that produced the candidate SQL and evaluate committees on that generator’s 82 items in two settings. The \emph{inclusive} setting retains all \(\binom{6}{k}\) committees, even when \(g\) is a member (self-judging is allowed). The \emph{exclusive} setting removes committees that contain \(g\), yielding \(\binom{5}{k}\) eligible committees for \(k \le 5\) and none for \(k=6\). 

\paragraph{Why include and exclude self-judging.}
Modern LLMs can meaningfully critique their \emph{own} outputs, and real deployments could possibly place a single model in both generator and reviewer roles. Measuring with self-judging therefore reflects a practical scenario in which the generator participates in a standing panel and can endorse or veto its own code. At the same time, self-judging may introduce optimism or style bias and can mask peer-review effects. Reporting results without self-judging isolates cross-model judgment and clarifies how well \emph{other} models vet a given generator’s outputs. Showing both variants quantifies the marginal value and potential risk of allowing a generator to sit in judgment of itself.

\section{Experiment Results}
\subsection{Baseline Text-to-SQL Accuracy of 15 Open Models}
\begin{figure}[htbp]
\centering
\includegraphics[width=\columnwidth]{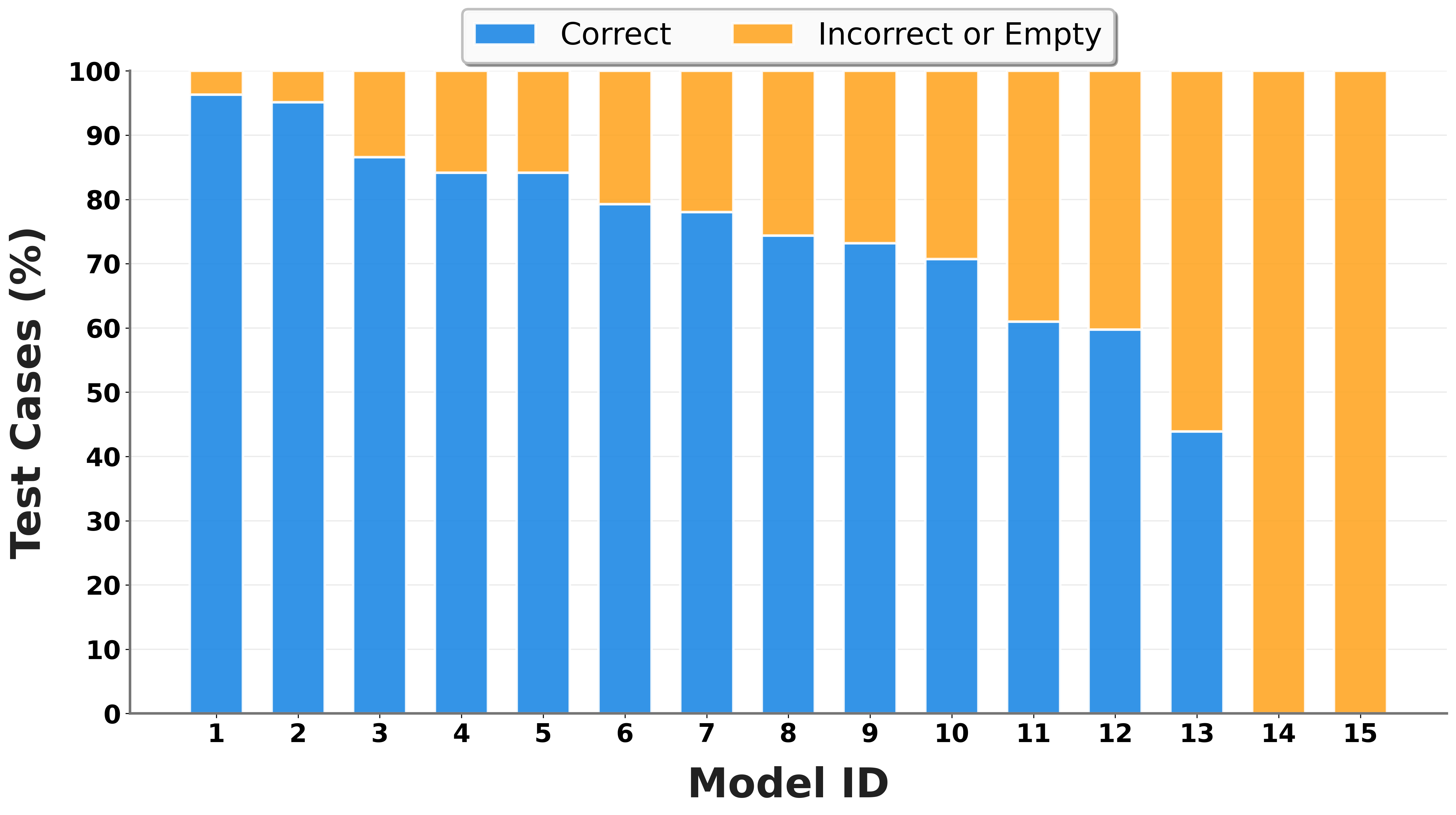}
\caption{Baseline Text-to-SQL Accuracy of 15 Open Models}
\label{baseline-1}
\end{figure}

Figure~\ref{baseline-1} summarizes how well each of the 15 open-source models performs as a \emph{coder} on our 82 text-to-SQL prompts. For every model we generate exactly one SQL answer at \texttt{temperature}=0 and label it using the execution-grounded protocol in Section~\ref{methods-1}. The blue segment shows the percentage of prompts for which the model produced a query whose results matched the ground truth on all 10 database instances (i.e., judged correct). The orange segment aggregates all other outcomes: syntactically invalid SQL, empty or non-SQL responses, and SQL whose execution results disagreed with the ground truth on at least one instance.

The figure shows a clear performance spread. A small group of models (IDs 1–5) succeeds on almost all prompts. These correspond to \texttt{gpt-oss:120b}, \texttt{gpt-oss:20b}, \texttt{codegemma:7b}, \texttt{phi4:14b}, and \texttt{gemma3:27b}, as mapped in Table~\ref{tab:model_specs}. This indicates that the strongest open models in our pool can reliably handle our schema, joins, and aggregation patterns. Performance then tapers through the middle tier (IDs 6–10), where models often produce plausible-looking SQL that nevertheless fails at least one execution check. The lowest tier (IDs 11–15) struggles to produce consistently executable or correct SQL under the same conditions. It is noteworthy that the models in this lowest tier are also the smallest in parameter count (all under 10B), suggesting that, for our workload, model capacity might be a major factor in text-to-SQL quality. 

\subsection{Aggregate Performance of Unanimous Committees}
We next study how conservative, unanimous committees behave when they are built from the strongest models in our pool. From the 15 models in Table~\ref{tab:model_specs}, we select the top six by single-attempt execution accuracy (see Figure~\ref{baseline-1}) and treat them as candidate \emph{judges}. As mentioned earlier, our argument is that it is more realistic to ask strong generators to review SQL than to rely on models that cannot themselves solve most of the tasks.

Using these six judges, we enumerate \emph{all} possible committees for sizes \(k=1,\dots,6\). Table~\ref{tab:judge-1} is the lookup table we use throughout the results section for the committees of judges. This lets us show large numbers of committees in the plots without repeating model names in every figure. For example, “Committee ID 1, \(k=3\)” always refers to models \((1,2,3)\), which are \texttt{gpt-oss:120b}, \texttt{gpt-oss:20b}, and \texttt{codegemma:7b} in Table~\ref{tab:model_specs}. “Committee ID 1, \(k=4\)” refers to \((1,2,3,4)\), and so on. All other committees are defined similarly by their ID and \(k\).

\begin{table}[ht]
\centering
\caption{Judge Committee Membership for Each Committee Size. Cells marked ``-'' are not applicable for that committee size.}
\label{tab:judge-1}
\renewcommand{\arraystretch}{0.7}
\setlength{\tabcolsep}{4pt}
\begin{tabular}{@{}lcccccc@{}}
\toprule
\textbf{Committee} & \textbf{k=1} & \textbf{k=2} 
& \textbf{k=3} & \textbf{k=4} 
& \textbf{k=5} & \textbf{k=6} \\
\textbf{Size} & & & & & & \\
\cmidrule{1-7}
\textbf{Committee} & \multicolumn{6}{c}{\textbf{Model IDs in each Committee of Judges}} \\
\textbf{ID} & & & & & & \\
\midrule

1  & (1) & (1,2) & (1,2,3) & (1,2,3,5) & (1,2,3,4,5) & (1,2,3,4,5,6) \\
2  & (2) & (1,4) & (1,2,5) & (1,2,3,4) & (1,2,3,5,6) & - \\
3  & (4) & (1,3) & (1,2,4) & (1,2,4,5) & (1,3,4,5,6) & - \\
4  & (5) & (1,5) & (1,3,4) & (1,3,4,5) & (1,2,3,4,6) & - \\
5  & (6) & (2,5) & (1,4,5) & (2,3,4,5) & (1,2,4,5,6) & -\\
6  & (3) & (2,3) & (1,3,5) & (1,2,5,6) & (2,3,4,5,6) & - \\
7  & - & (2,4) & (2,3,5) & (1,2,3,6) & - & - \\
8  & - & (2,6) & (2,4,5) & (1,4,5,6) & - & - \\
9  & - & (1,6) & (2,3,4) & (1,3,4,6) & - & - \\
10 & - & (3,4) & (1,2,6) & (1,2,4,6) & - & - \\
11 & - & (4,5) & (1,4,6) & (2,3,5,6) & - & - \\
12 & - & (4,6) & (2,3,6) & (1,3,5,6) & - & - \\
13 & - & (3,5) & (2,5,6) & (2,3,4,6) & - & - \\
14 & - & (5,6) & (1,3,6) & (2,4,5,6) & - & - \\
15 & - & (3,6) & (1,5,6) & (3,4,5,6) & - & - \\
16 & - & - & (2,4,6) & - & - & - \\
17 & - & - & (3,4,5) & - & - & - \\
18 & - & - & (3,4,6) & - & - & - \\
19 & - & - & (4,5,6) & - & - & - \\
20 & - & - & (3,5,6) & - & - & - \\
\bottomrule
\end{tabular}
\end{table}

\begin{figure}[htbp]
\centering
\includegraphics[width=\columnwidth]{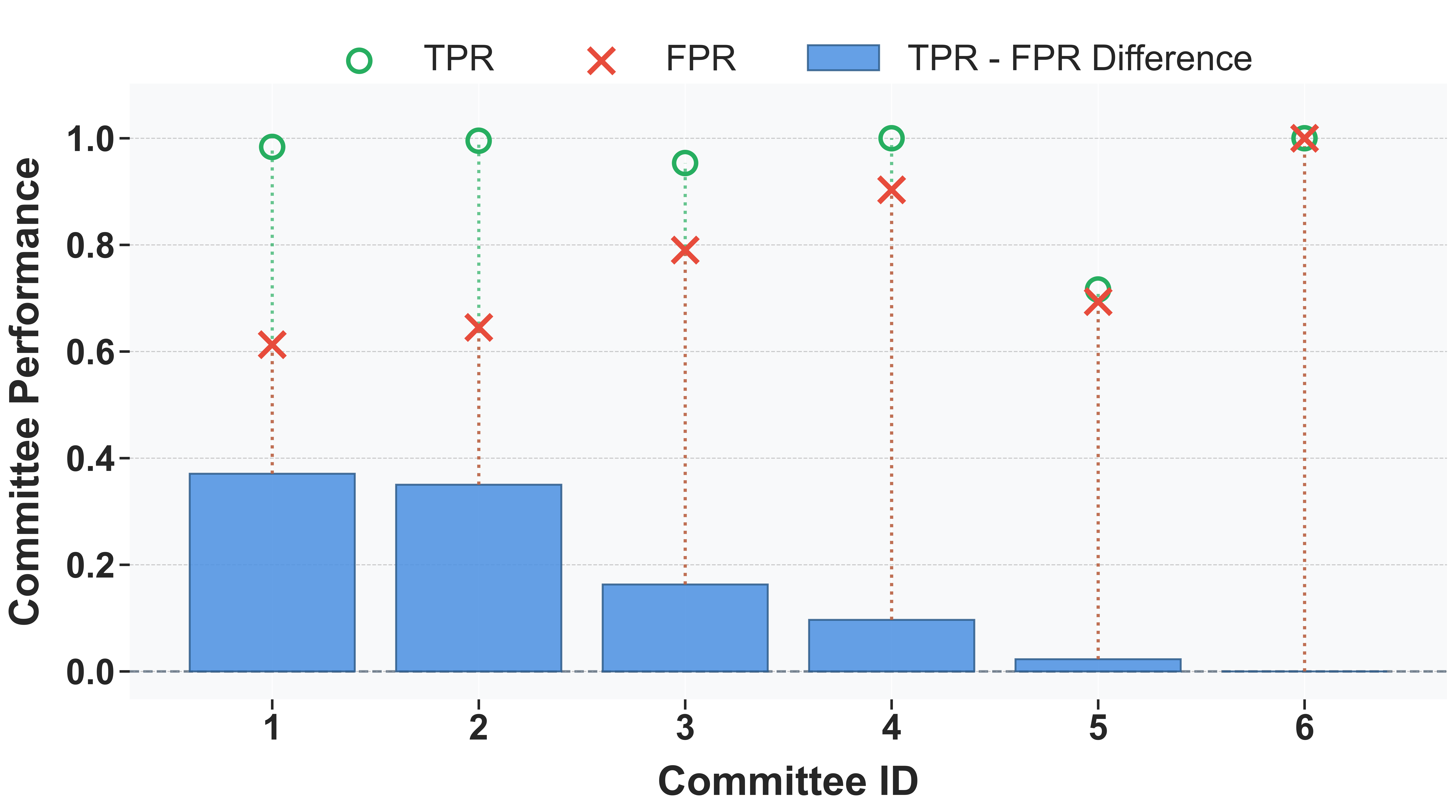}
\caption{Aggregate performance of single-model committees (\(k=1\)) over all 492 queries. Bars show Youden's \(J = \mathrm{TPR} - \mathrm{FPR}\); green circles mark TPR; red crosses mark FPR. Committees are ordered by \(J\).}
\label{fig-k-1}
\end{figure}

\begin{figure}[htbp]
\centering
\includegraphics[width=\columnwidth]{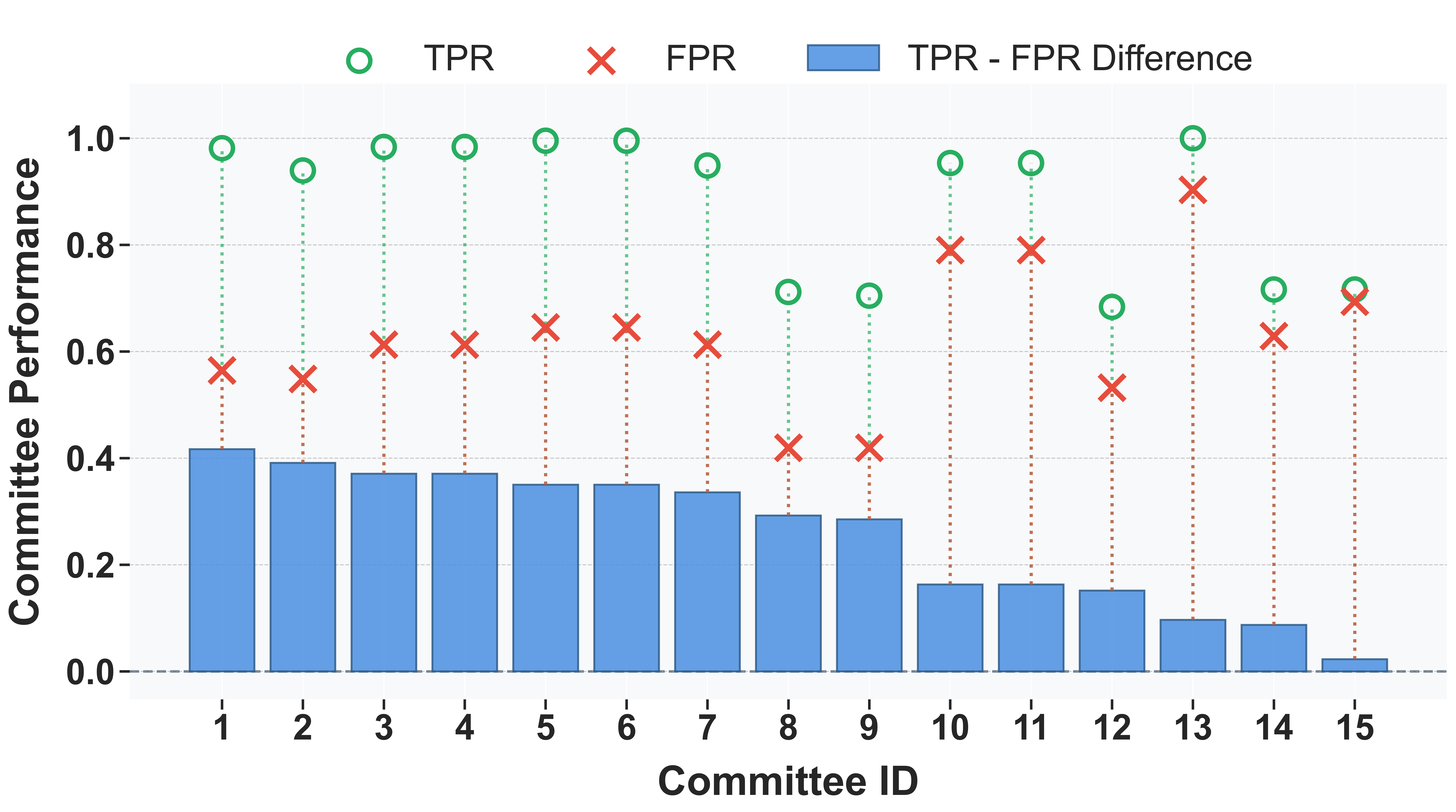}
\caption{Aggregate performance of all \(\binom{6}{2}=15\) two-model committees (\(k=2\)) under the unanimous rule, evaluated on 492 queries. Bars show Youden's \(J\); green circles mark TPR; red crosses mark FPR. Committees are ordered by \(J\).}
\label{fig-k-2}
\end{figure}

\begin{figure}[htbp]
\centering
\includegraphics[width=\columnwidth]{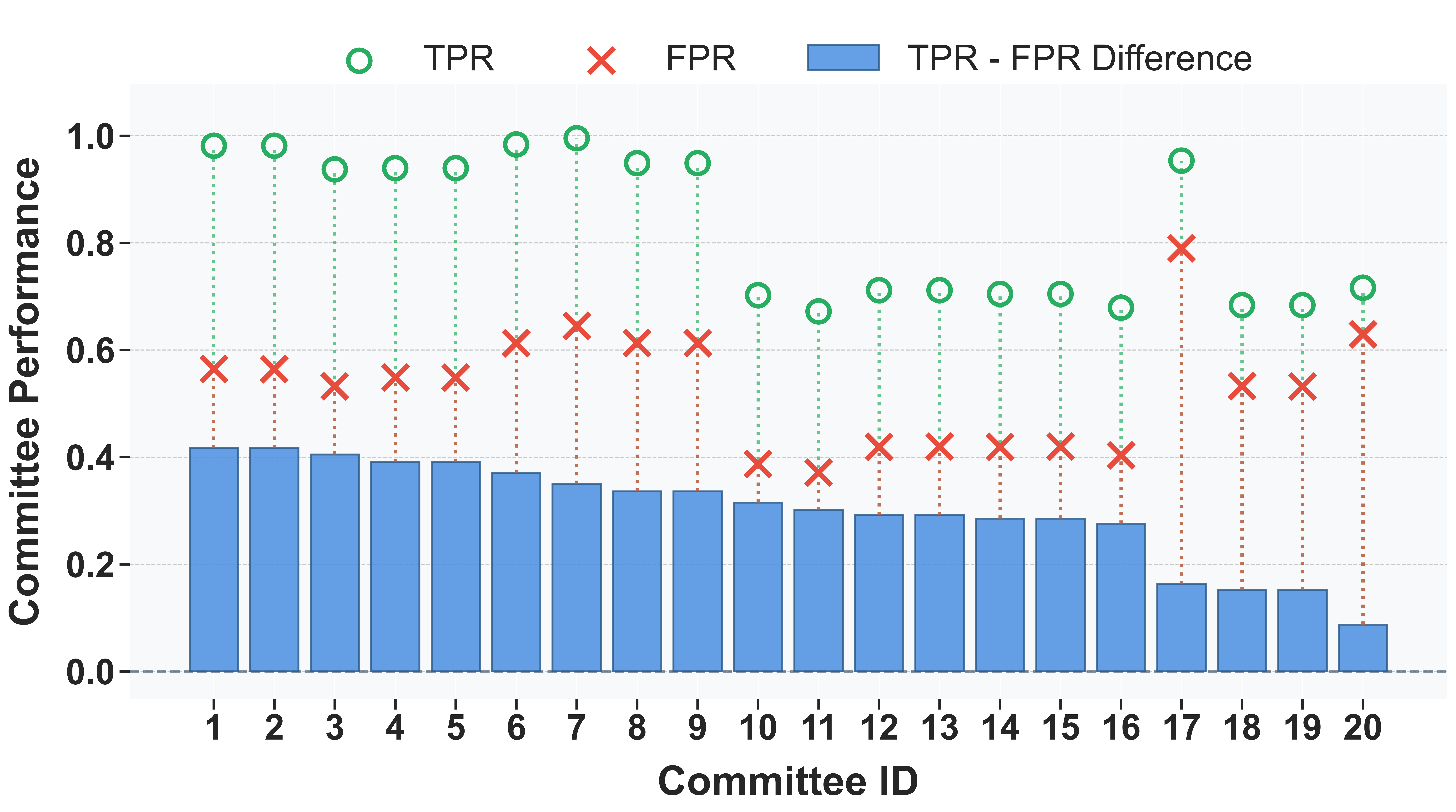}
\caption{Aggregate performance of all \(\binom{6}{3}=20\) three-model committees (\(k=3\)) under the unanimous rule, evaluated on 492 queries. Bars show Youden's \(J\); green circles mark TPR; red crosses mark FPR. Committees are ordered by \(J\).}
\label{fig-k-3}
\end{figure}

\begin{figure}[htbp]
\centering
\includegraphics[width=\columnwidth]{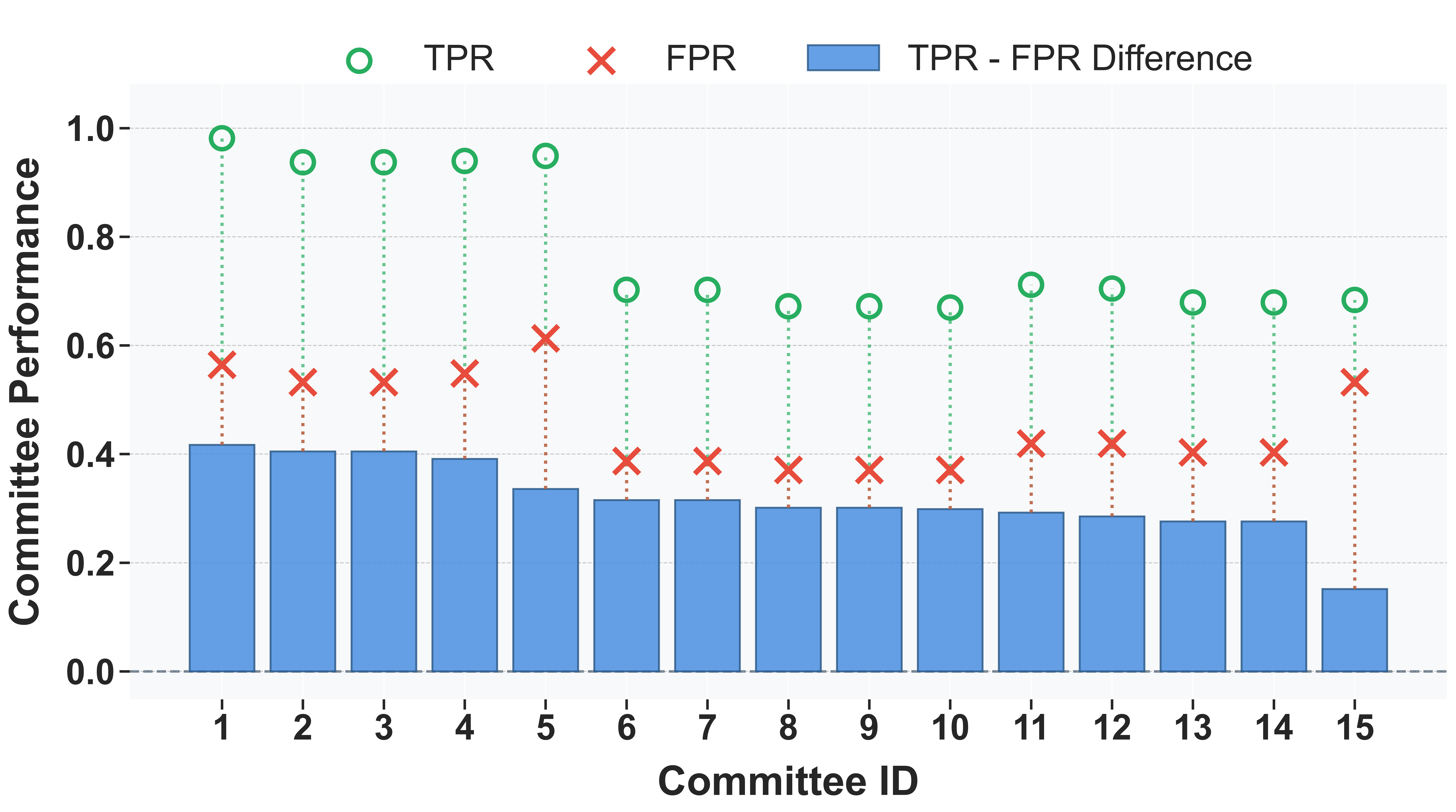}
\caption{Aggregate performance of all \(\binom{6}{4}=15\) four-model committees (\(k=4\)) under the unanimous rule, evaluated on 492 queries. Bars show Youden's \(J\); green circles mark TPR; red crosses mark FPR. Committees are ordered by \(J\).}
\label{fig-k-4}
\end{figure}

\begin{figure}[htbp]
\centering
\includegraphics[width=\columnwidth]{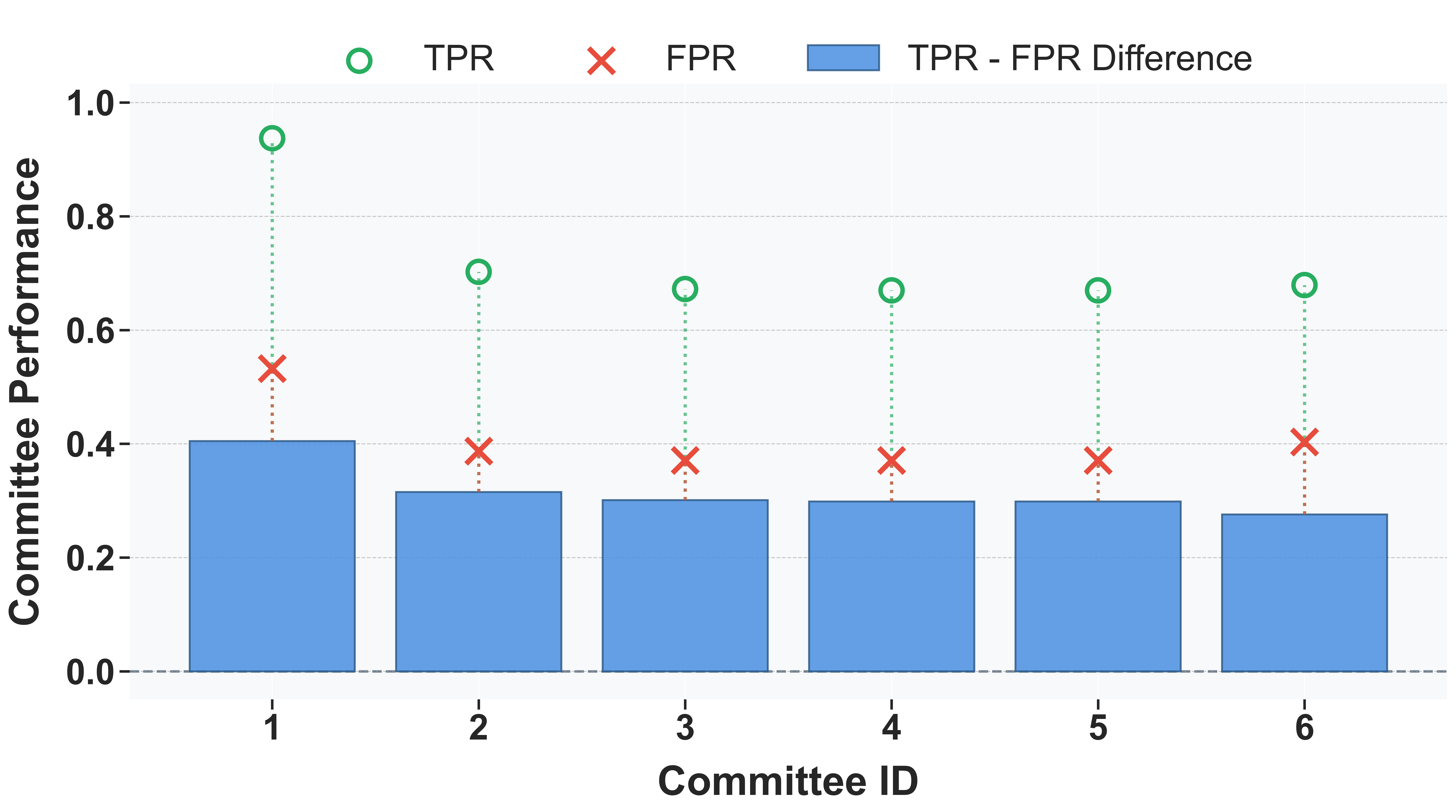}
\caption{Aggregate performance of all \(\binom{6}{5}=6\) five-model committees (\(k=5\)) under the unanimous rule, evaluated on 492 queries. Bars show Youden's \(J\); green circles mark TPR; red crosses mark FPR. Committees are ordered by \(J\). The unique \(k=6\) committee is discussed in text.}
\label{fig-k-5}
\end{figure}

Figures~\ref{fig-k-1}--\ref{fig-k-5} compare all committees of sizes \(k=1\) through \(k=5\) on the full set of 492 candidate queries (82 prompts \(\times\) 6 generators). Each plot uses the same visual encoding: the bar height is Youden’s \(J = \mathrm{TPR} - \mathrm{FPR}\); green circles show the true–positive rate (share of truly correct queries that were accepted); red crosses show the false–positive rate (share of truly incorrect queries that were accepted). Higher \(J\) indicates better discrimination at this unanimous operating point, but two committees with similar \(J\) can still make different trade-offs between TPR and FPR, which is why we show all three values. Within each figure, committees are sorted by \(J\).

Figure~\ref{fig-k-1} shows that for single-model committees (\(k=1\)), the best value of \(J\) sits just below 0.4 (Committee IDs 1 and 2). Individual judges differ markedly: some reach very high TPR (close to 1.0) but at the cost of high FPR (above 0.6), meaning they are willing to pass many incorrect queries. Others are more conservative but also accept far fewer correct queries, which lowers \(J\). This confirms that “use one LLM as the reviewer” does not give a stable or uniformly safe operating point.

Moving to unanimous pairs (Figure~\ref{fig-k-2}) and unanimous triplets (Figure~\ref{fig-k-3}) improves the picture slightly. The best committees now keep FPR below about 0.6 while retaining TPR just under 1.0, so \(J\) nudges upward. In other words, unanimity among two or three strong judges is already enough to filter out a substantial fraction of bad SQL while still letting most good SQL through. The spread across committees is still visible, which tells us that \emph{which} models you group together matters, not just the value of \(k\).

For four- and five-member unanimous committees (Figures~\ref{fig-k-4} and \ref{fig-k-5}), FPR becomes low and fairly uniform, but TPR also compresses downward. This is the natural effect of unanimity: every extra judge is another chance to veto. The result is a flatter set of \(J\) bars, because most committees are now very strict. This is a good operating region for teams that care more about avoiding false accepts than about automatic coverage. There is only one unanimous committee of size six (not plotted, since it has a single data point); it continues the same trend.

Across all values of \(k\), committees containing the two strongest GPT-style models (\texttt{gpt-oss:120b} and \texttt{gpt-oss:20b}) often appear among the top-\(J\) groups. By contrast, committees that included the weaker generators from our pool (e.g., \texttt{gemma3:27b} or \texttt{codellama:34b}, see Table~\ref{tab:model_specs}) were more likely to show up toward the bottom of the ranking for that \(k\) (for example, Committees~14 and~15 for \(k=2\), several of the rightmost committees for \(k=3\), and the tail for \(k=4\)). This pattern reinforces the design choice to build judge pools from the strongest available generators.

%  Per model
\subsection{Performance of Unanimous Committees per Generator}
The aggregate results in the previous subsection assume that a committee is asked to judge SQL coming from \emph{all} six generators at once. In practice, however, many deployments look more like the opposite: an organization standardizes on one (or a very small number of) text-to-SQL models, and every candidate query from that model is sent to a panel for review. In that setting, what matters is not “How good is this committee on 492 mixed queries?” but rather “How good is this committee \emph{on the outputs of generator \(g\)}?” This is why we break results down by generator.

\begin{figure}[htbp]
\centering
\includegraphics[width=\columnwidth]{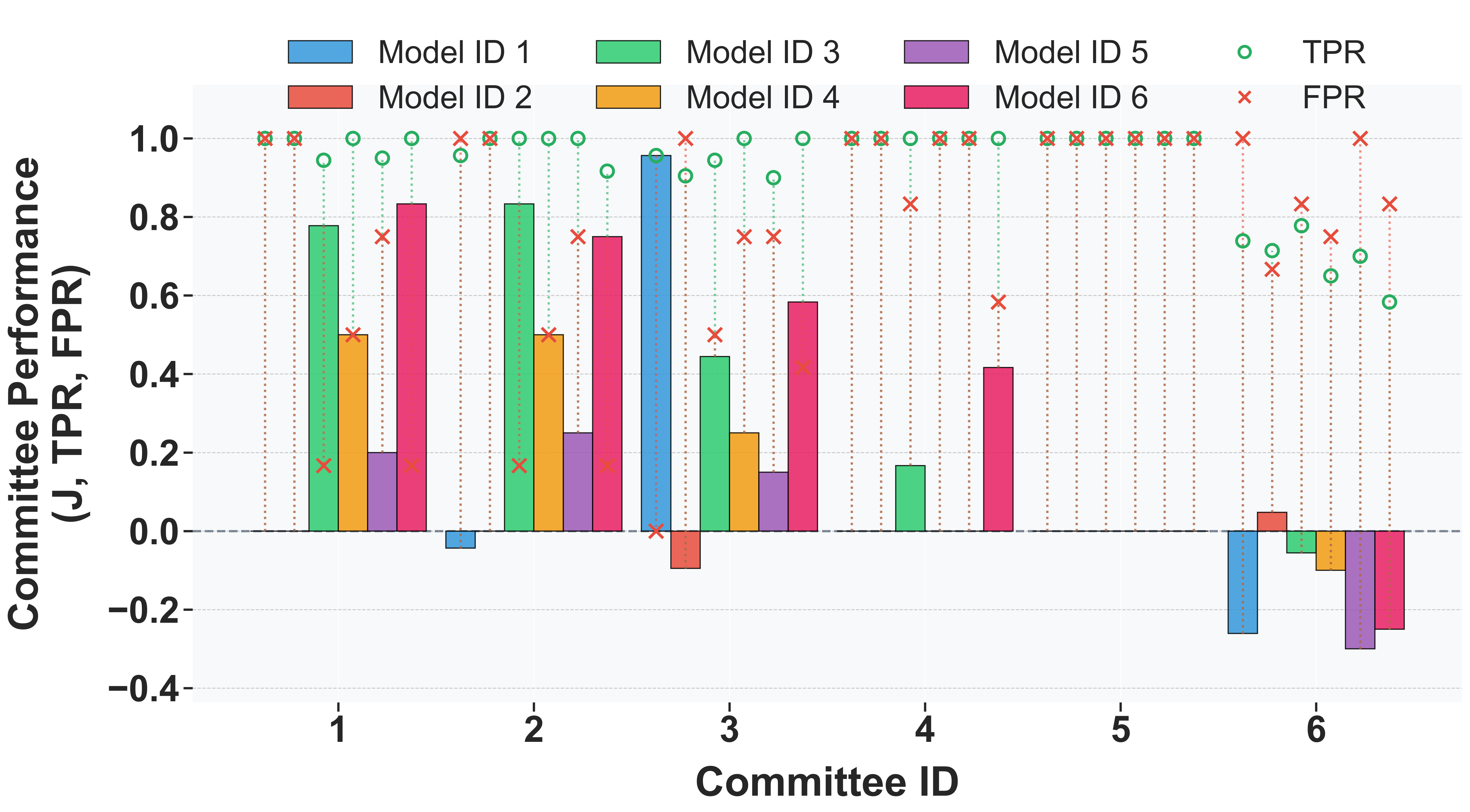}
\caption{Per-generator performance of single-model committees (\(k=1\)). Each committee (x-axis) judges the 6 generators separately; bars show Youden's \(J\) for each generator, colored by generator ID, while green circles and red crosses show TPR and FPR respectively. Gaps within a committee indicate \(J=0\) for that generator.}
\label{fig:pergen-k1}
\end{figure}

\begin{figure}[htbp]
\centering
\includegraphics[width=\columnwidth]{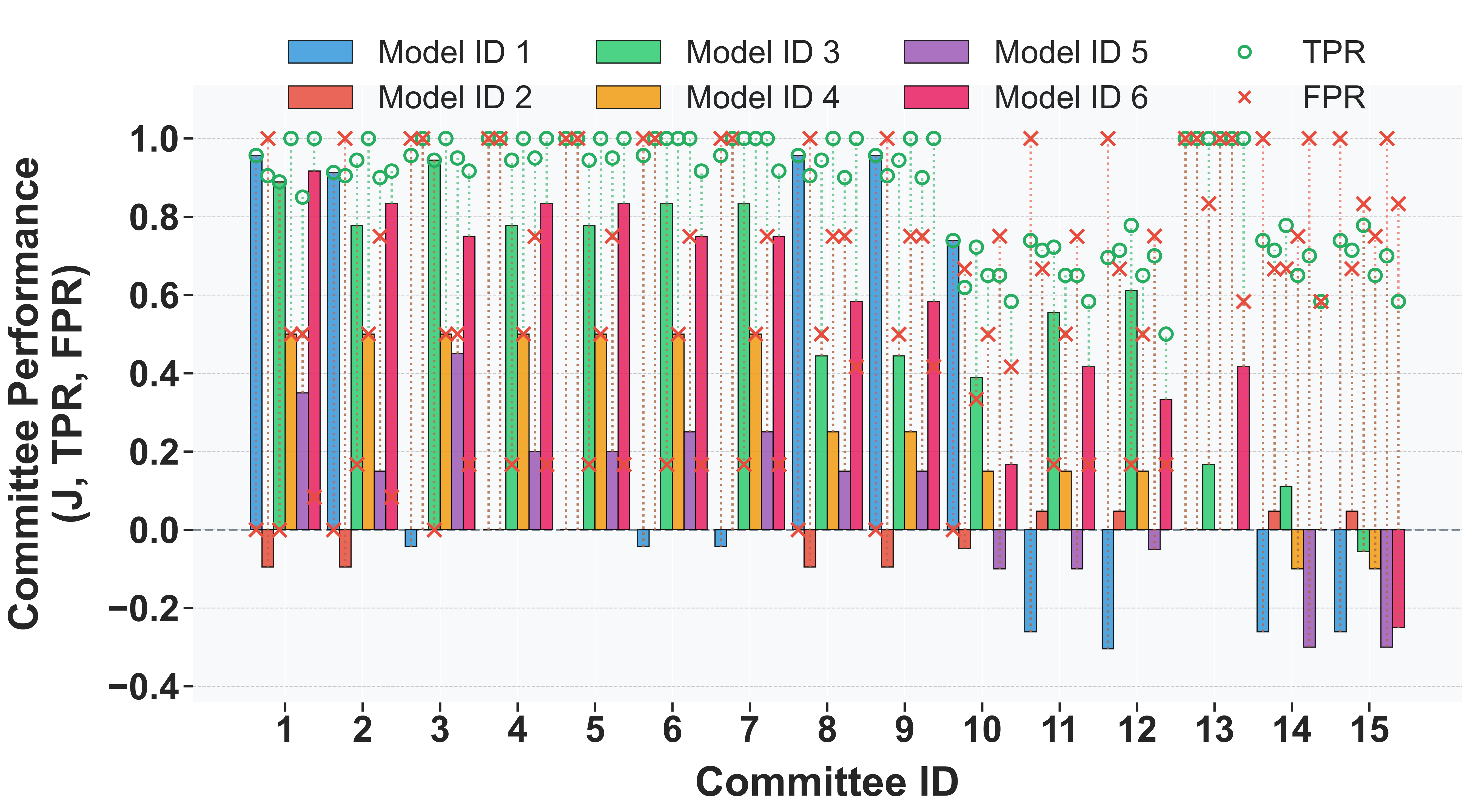}
\caption{Per-generator performance of all \(\binom{6}{2}=15\) two-model committees (\(k=2\)) under the unanimous rule. Each committee (x-axis) judges the 6 generators separately; bars show Youden's \(J\) for each generator, colored by generator ID, while green circles and red crosses show TPR and FPR respectively. Gaps within a committee indicate \(J=0\) for that generator.}
\label{fig:pergen-k2}
\end{figure}

\begin{figure}[htbp]
\centering
\includegraphics[width=\columnwidth]{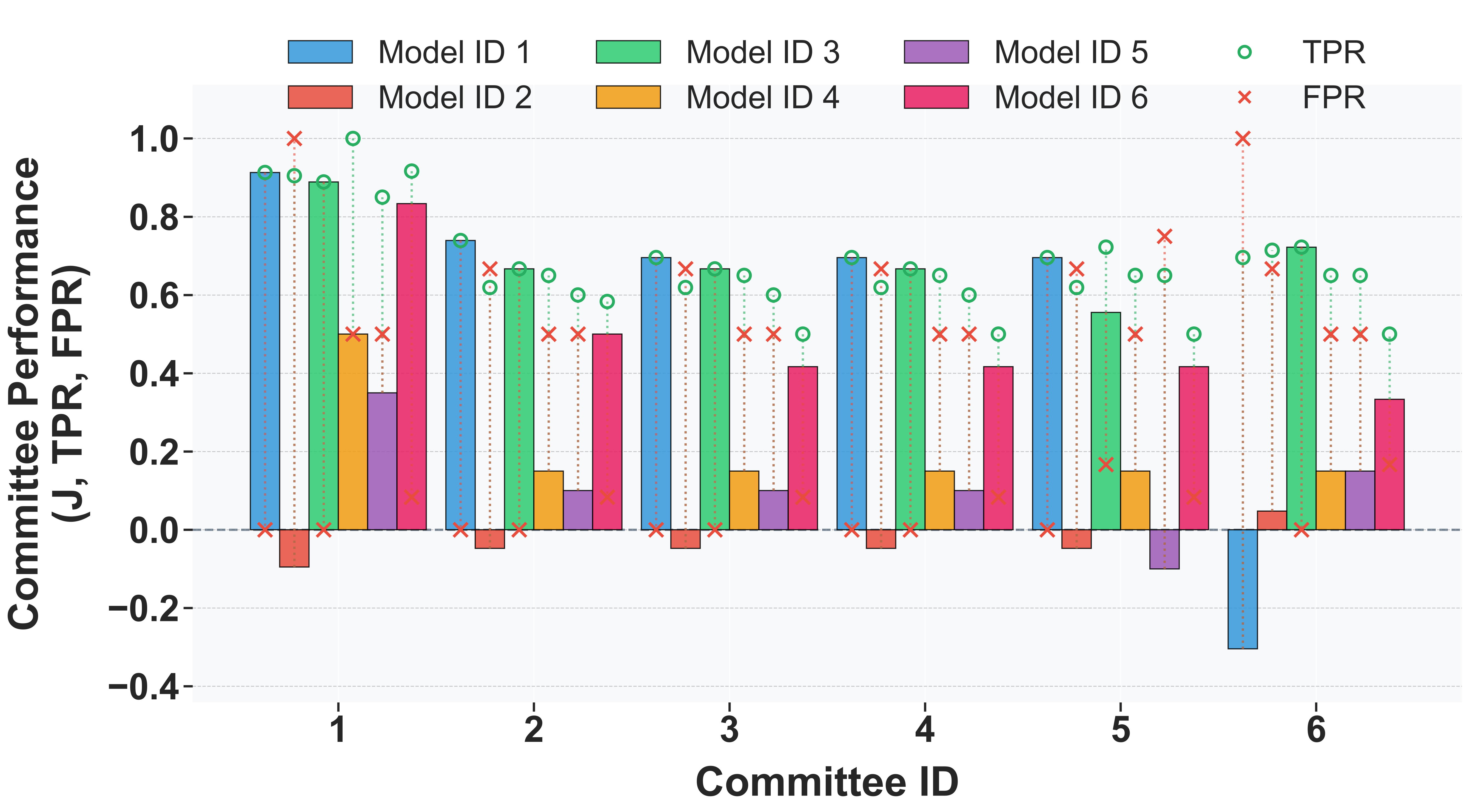}
\caption{Per-generator performance of the six five-model committees (\(k=5\)) under the unanimous rule. Each committee (x-axis) judges the 6 generators separately; bars show Youden's \(J\) for each generator, colored by generator ID, while green circles and red crosses show TPR and FPR respectively. Gaps within a committee indicate \(J=0\) for that generator.}
\label{fig:pergen-k5}
\end{figure}

Figures~\ref{fig:pergen-k1}--\ref{fig:pergen-k5} show this per-generator view for \(k=1\), \(k=2\), and \(k=5\); the other \(k\) values tell the same story and are omitted for space. For a fixed \(k\), the x-axis lists the committees in the same order as before, and each committee is aligned to a cluster of six bars, one bar per generator being judged (color-coded by model ID). Bars plot Youden’s \(J\); overlaid green circles and red crosses plot TPR and FPR so that we can see \emph{how} each \(J\) was achieved. A missing bar in a cluster means that generator’s \(J\) was zero for that committee.

\textbf{Single-model judges (\(k=1\)).}
Figure~\ref{fig:pergen-k1} already shows large variation. The best judge until this point (Committee~1, \texttt{gpt-oss:120b}) gives very high \(J\) to some \emph{other} generators. For example, it scores Model~6 (\texttt{codellama:34b}) above 0.8 and Model~3 (\texttt{codegemma:7b}) just under 0.8, meaning cross-model judgment can be strong even without self-judging. At the same time, Committee~6 (the weakest single judge) produces negative or near-zero \(J\) for almost every generator. This matches what we saw in aggregate and confirms that some models are simply unreliable reviewers.

\textbf{Self-judging is not automatically good.}
These plots also show that “let the model review itself” is \emph{not} a guaranteed win. In Figure~\ref{fig:pergen-k1}, Model~6 judging its own SQL yields \(J\approx 0\): it neither accepts the right things consistently nor rejects the wrong things consistently. In contrast, when strong generators sit on the panel, self-judging tends to look very good (high TPR, low-ish FPR), which is what we later see for the GPT-style models. So the usefulness of self-judging is \emph{model-dependent}: it seems to help for the top generators but not for weaker ones.

\textbf{Two-model unanimous committees (\(k=2\)).}
In Figure~\ref{fig:pergen-k2}, the pattern sharpens. Some two-model panels that contain a strong generator give that generator a \(J\) close to 1.0, while giving much lower \(J\) to other generators. Other panels, especially those that include weaker models, still produce low or even negative \(J\) for certain generators. This tells us that simply adding “one more model” to the panel is not enough; the \emph{composition} of the panel relative to the generator being judged matters.

\textbf{Five-model unanimous committees (\(k=5\)).}
Figure~\ref{fig:pergen-k5} shows the most conservative setting. With five out of six models on every panel, FPRs tend to be modest across generators, but \(J\) does not collapse to a single value: some generators still get much better treatment from some panels than others. This is an important practical point: even with a large, strict committee, a team that standardizes on one generator should still measure judgment quality \emph{for that generator} rather than rely only on aggregate committee numbers.

Overall, these per-generator plots make three points. First, the per-generator view is closer to deployment reality, where one model produces most of the SQL. Second, self-judging can be excellent for the strongest generators but can be ineffective or even harmful for weaker ones, so results should be reported both with and without self-judging. Third, committee composition still matters at fixed \(k\): some panels consistently produce high \(J\) for a given generator, while others do not, which justifies the exhaustive committee sweep in the previous subsection.

% \begin{figure}[htbp]
% \centering
% \includegraphics[width=\columnwidth]{Figures/comm_3_per_model_tpr_fpr_diff.png}
% \caption{TPR Distribution by Committee Size}
% \label{fig}
% \end{figure}

% \begin{figure*}[htbp]
% \centering
% \includegraphics[width=\columnwidth]{Figures/comm_4_per_model_tpr_fpr_diff.png}
% \caption{TPR Distribution by Committee Size}
% \label{fig}
% \end{figure*}

% \begin{figure}[htbp]
% \centering
% \includegraphics[width=\columnwidth]{Figures/boxplot_youden_j.png}
% \caption{TPR Distribution by Committee Size}
% \label{fig}
% \end{figure}

% \begin{figure*}[htbp]
% \centering
% \includegraphics[width=\textwidth]{Figures/flowchart.pdf}
% \caption{Fow Diagram }
% \label{fig}
% \end{figure*}

\section{Discussion and Conclusions}
Our findings suggest a practical recipe for teams adopting “vibe coding” to SQL: prefer a small, unanimous panel composed of strong models over a single judge. Unanimity delivers a predictable safety gain (i.e., lower FPR without collapsing TPR), especially at moderate values of k. However, committees are not interchangeable. The same k can yield very different operating behavior depending on which models are used, and the best panel in aggregate may not be best for your chosen generator. This argues for per-generator validation: if one model produces most of your SQL, measure judgment quality on that source specifically. Self-judging is not a universal win; it benefits top generators but can be ineffective for weaker ones, so deployments should test both inclusive (with self-judging) and exclusive configurations.

Limitations point to clear next steps. We study a single schema family on MySQL with 82 prompts, temperature-0 decoding, and an execution-based label that, while reproducible, abstracts away human intent nuances. We evaluate unanimous voting but not alternative rules (e.g., majority or cost-sensitive thresholds), and we do not optimize for latency or cost. Future work includes expanding schemas and DBMSs, adding harder compositional queries, exploring adaptive or weighted committees learned from per-judge calibration, and integrating cost/risk trade-offs to select panels and thresholds automatically. Despite these bounds, our results provide an actionable middle ground between single-judge heuristics and full human review. %: small, unanimous committees of strong open models can materially reduce false accepts while keeping useful coverage, and their behavior can be audited and tuned to the generator actually in use.

\bibliographystyle{IEEEtran}
\bibliography{main}

\end{document}